\documentclass[prb,twocolumn,superscriptaddress,showpacs]{revtex4}
\usepackage{amsmath,amssymb,graphicx,hyperref}

\begin{document}
\title{
Entangled photons from a strongly coupled quantum dot-cavity system}

\author{Robert Johne}
\affiliation{LASMEA, UMR 6602 CNRS, Universit\'{e} Blaise Pascal,
24 av.~des Landais, 63177 Aubi\`ere, France}

\author{Nikolay A. Gippius}
\affiliation{LASMEA, UMR 6602 CNRS, Universit\'{e} Blaise Pascal,
24 av.~des Landais, 63177 Aubi\`ere, France}
\affiliation{A.M. Prokhorov General Physics Institute, RAS, 119991, Moscow, Russia}

\author{Guillaume Malpuech}
\affiliation{LASMEA, UMR 6602 CNRS, Universit\'{e} Blaise Pascal,
24 av.~des Landais, 63177 Aubi\`ere, France}

\date{\today}

\begin{abstract}

A quantum dot strongly coupled to a photonic crystal has been recently proposed as a source of entangled photon pairs [R. Johne et al., Phys. Rev. Lett. 100, 240404 (2008)]. The biexction decay via intermediate polariton states can be used to overcome the natural splitting between the exciton states coupled to the horizontally and vertically polarized light modes, so that high degrees of entanglement can be expected. We investigate theoretically the features of realistic dot-cavity systems, including the effect of the different oscillator strength of excitons resonances coupled to the different polarizations of light. We show that in this case, an independent adjustment of the cavity resonances is needed in order to keep a high entanglement degree. We also consider the case when the biexciton-exciton transition is also strongly coupled to a cavity mode. We show that a very fast emission rate can be achieved allowing the repetition rates in the THz range. Such fast emission should however be paid for by a very complex tuning of the many strongly coupled resonances involved and by a loss of quantum efficiency. Altogether a strongly coupled dot-cavity system seems to be very promising as a source of entangled photon pairs.

\end{abstract}


\maketitle

\section{
Introduction}

The strong light-matter coupling in a solid state system has first been observed for bulk polarons \cite{Huang} and exciton-polaritons.\cite{Hopfield} However, their key properties, such as the dispersion, could not be tailored at will. An important step forward was the observation of strong coupling in a planar semiconductor microcavity with a quantum well in 1992. \cite{Weisbuch}  In such a system the properties of the new 2D excitations named cavity exciton-polaritons (polaritons) \cite{Microcavities} can be controlled in a wide range by changing the properties of the cavity or the quantum well. 

Due to the strong improvement of the technology and fabrication techniques it was later possible to realize a true solid state implementation
of the model interacting light-matter system, that is, a system where a single atomic resonance strongly interacts with the confined photonic mode of a cavity. In solid state systems, the atomic resonance is replaced by the excitonic resonance of a quantum dot (QD) which is embedded in 
an optical microcavity e.g. a micro-pillar,\cite{Reithmaier2004} a micro disk,\cite{Peter2005} or a photonic crystal. The analogy is based on the assumption that the optical response of a semiconductor QDs is truly atomic like, so that the dot can absorb a single photon at a given energy which is not evident from a theoretical point of view. Especially when the QDs used to observe the strong coupling are the large fluctuations on the interfaces of a quantum well. \cite{LaussyQD} The demonstration that a real "quantum regime" can be achieved 
came actually in 2007 with the report of photon counting experiments demonstrating that the measured emission doublet, signature of the strong coupling, was occurring when no more than a single photon is present within the cavity at a given time. \cite{Hennessy2007}
 This progress has been made possible because of the development of the photonic crystal technology, which allows to achieve incredibly high quality factor, to tune energies of the photonic modes, and also to place the QDs at a well defined position within the microcavity.\cite{Hennessy2006, Vuckovic, Hennessy2007, Badolato, Winger}

Another fundamental effect of quantum mechanics in the focus of research since the development of quantum computing and quantum communication \cite{Bennet1984, Bennet1993} is the quantum entanglement, first discussed by Einstein, Podolsky, and Rosen.\cite{Einstein} In that context
 QDs have drawn a strong attention when it has been proposed to realize solid state entangled photon pair emitters based on biexciton decay. \cite{Benson2000} The ideal decay paths are illustrated in Fig.1(a). The biexciton decays emitting either first a $\sigma_{+}$ and second a $\sigma_{-}$ photon or vice versa, and the photons are fully polarization-entangled. Unfortunately, this proposal turned out to be hard to implement mainly because the intermediate exciton states of a typical QDs are not degenerate due to the anisotropic electron-hole interaction \cite{Gammon1996,Kulakovskii1999}. This interaction couples degenerate exciton states which split into two resonances coupled to two orthogonal linear polarizations called horizontal (H) and vertical (V), respectively. The resulting photons for the two decay channels are therefore distinguishable. The quantum correlations become hidden in time integrated measurements because a QD with split intermediate exciton levels emits photons into a time-evolving entangled state.\cite{Stevenson2008}

Several proposals have been made to overcome this splitting of the exciton lines \cite {Vogel2007,Gerardot2007} and several reports on the generation of entangled photon pairs have been published.\cite{Stevenson,Akopian2006,Hafenbrak,Young2007} 
From our side, we proposed recently to use the new possibilities of engineering of QDs optical response offered by the achievement of the strong coupling regime.\cite{Johne2008} We have shown that it is possible to tune the energy of the mixed exciton-photon states in order to recover the  degeneracy and achieve a high degree of entanglement for the photon pairs emitted by the biexciton cascade. The only requirement is that the photon levels polarized along H and V direction should be split by an energy of a few tens of meV, which is exactly what generally happens for the cavity modes in photonic crystals. The advantage of this scheme is that it does not require a very precise set of structural parameters,
and seems achievable independently of the exact values of the splitting between the 
bare exciton if the conditions for the achievement of the strong coupling regime are met. This avoids in principle to discover among thousands of QDs the unique one which will show degenerate exciton levels. Another advantage is that the intermediate polariton states are emitting photons within 10-20 ps typically against 1 ns for a bare exciton level, which is a very good protection against all possible dephasing processes. In this scheme, one should understand that the important aspect is not that the polariton state itself is an entangled exciton-photon state. This aspect is useless in a solid state system since there is no clear way to measure directly the exciton state. The key property which is used here is the possibility to modify the energy position of an electronic state by strongly coupling this state to the optical resonance of a photonic crystal.

On the other hand we have neglected \cite{Johne2008} some important specificities of optical response of real QDs, the main one being
the fact that the exciton resonances coupled to the H and V polarized modes do not show the same oscillator strength,\cite{Favero2005} which at first 
glance may seem highly detrimental for the effect we propose. Another thing neglected was the possible coupling of the biexciton-exciton transition  with some optical resonances of the photonic crystal. 

The goal of the present paper is to include these important effects in the theory in order to understand their consequences
on the entanglement degree of the emitted photon pairs. We believe that careful accounting of these specificities
would be of a great help for the experimental groups who may desire to realize the scheme we propose.
 In section II we present the analytical description of the system and remind briefly the favorable scheme presented in Ref\cite{Johne2008}. In section III we analyze the impact of different oscillator strength on the degree of entanglement and develop new working configurations of a strongly coupled dot-cavity system as an emitter of entangled photons. Section V takes into account spectroscopic filtering and section V discusses the possible impact of additional resonances close to the biexciton-exciton transition energy. Finally, a summary and conclusions are presented in the last part VI.

\section{Initial scheme and analytical description}

\begin{figure}
\begin{center}
\includegraphics[width=1.0\linewidth]{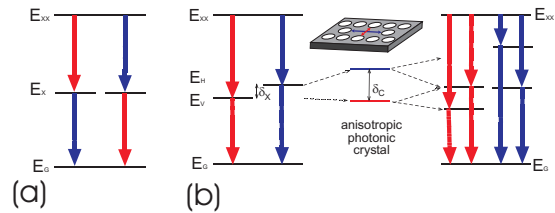}
\caption{ \label{fig1} (a) ideal decay path of a quantum dot; (b) biexciton decay of a real quantum dot (left), photonic crystal resonances (middle) and resulting decay scheme for the quantum dot embedded in the photonic crystal in the strong coupling regime (right). The blue arrows correspond to horizontal polarization and the red arrows correspond to vertical polarization.}
\end{center}
\end{figure}

The biexciton decay scheme for an ideal quantum dot is shown in Fig.\ref{fig1}(a). The intermediate exciton states are degenerate and they couple to circularly polarized light. In a real QD Fig.\ref{fig1}(b, left part) the exciton resonances coupled to H and V polarized light modes are typically split by an energy $\delta_X$. We consider that such a QD is embedded within a photonic crystal, slightly anisotropic, which shows two confined optical modes polarized along H and V directions and split by a quantity $\delta_C$  (see the middle of Fig.\ref{fig1}(b)). 
Each of the two non-degenerate exciton states strongly couples to one resonance of the photonic crystal with either vertical (red) or horizontal (blue) polarization, respectively. This coupling gives rise to two polariton doublets polarized H and V. The resulting decay paths of the strongly coupled dot-cavity system can be seen on the Fig.\ref{fig1}(b, right part).  

There are now two possible decay channels for each polarization using either lower or upper polariton state. In Ref.\cite{Johne2008}
we have shown that for any constant values of $\delta_X$ and $\delta_C$, the adjustment of the energy detuning between the group
of exciton resonances and the group of photon resonances allows to make one polariton state with horizontal
polarization degenerate with one polariton state with vertical polarization. This alignment makes the two possible decay paths of the biexciton 
using these two intermediate states distinguishable only by their polarization which results in the generation of entangled photon 
pairs showing a maximum degree of entanglement.

The energy of the polariton states $E_{\pm}^{H,V}$ can be calculated using \cite{Microcavities}

\begin{equation} 
\label{eq1} {
 E_{\pm}^{H,V}=\frac{E_{C}^{H,V}+E_{X}^{H,V}}{2}\pm\frac{1}{2}\sqrt{(E_{C}^{H,V}-E_{X}^{H,V})^2+4\hbar^2\Omega_{H,V}^{2},}
 }
 \end{equation}
where H and V indicate the different polarizations, $E_{C}^{H,V}$ are the cavity resonances, $E_{X}^{H,V}$ are the exciton energies, and $\Omega_{H,V}$ are the values of Rabi splitting, proportional to the exciton oscillator strength which we now assume to be different for H and V polarized modes. It follows directly from Eq.(\ref{eq1}) that the energies of the intermediate polariton states can be tuned by changing the energy of the photonic resonances.

To describe the full decay scheme analytically we write the two-photon wavefunction in the following way neglecting cross polarization terms:
\begin{eqnarray}
\label{eq2} 
\left|\Psi\right\rangle=\left(\alpha_{LP}\left|p_H^{LP}\right\rangle+\alpha_{UP}\left|p_H^{UP}\right\rangle\right)\left|HH\right\rangle+\\ \nonumber
+\left(\beta_{LP}\left|p_V^{LP}\right\rangle+\beta_{UP}\left|p_V^{UP}\right\rangle\right)\left|VV\right\rangle,
\end{eqnarray}
where we extract the coordinate part $\left|p_{H(V)}^{LP(UP)}\right\rangle$ from the polarization part of the wavefunction $\left|HH\right\rangle (\left|VV\right\rangle)$. The amplitudes $\alpha$ and $\beta$ are the weights for the possible decay paths satisfying
\begin{equation}
\label{eq3} {
|\alpha_{LP}|^2+|\alpha_{UP}|^2+|\beta_{LP}|^2+|\beta_{UP}|^2=1.}
\end{equation}

After tracing out over all possible degrees of $|p\rangle$, the corresponding 2 photon density matrix reads

\begin{equation}
\label{eq4} {
\rho=\left|\Psi\right\rangle\left\langle \Psi\right|=\begin{pmatrix} \left|\alpha_{LP}\right|^2+\left|\alpha_{UP}\right|^2 & 0 & 0 & \gamma \\ 0 & 0 & 0 & 0 \\ 0 & 0 & 0 &0 \\
\gamma^* &0 &0 & \left|\beta_{LP}\right|^2+\left|\beta_{UP}\right|^2 \end{pmatrix},}
\end{equation}
where
\begin{equation}
\label{eq5}{
\gamma=\alpha_{LP}\beta_{UP}^*\left\langle p_H^{LP}|p_V^{UP}\right\rangle.}
\end{equation}

We select only the degenerate intermediate states using spectral windows, represented by a projection $P$, around the biexciton emission energy $E_{XX}$ and the polariton energy $E_P$. 
We can write the off-diagonal element in the following way:

\begin{equation}
\label{eq6}{
\gamma'=\frac{\alpha_{LP}\beta_{UP}^*\left\langle p_H^{LP}|P|p_V^{UP}\right\rangle}{\left|\alpha_{LP}\right|^2\left|\left\langle p_H^{LP}|P|p_V^{LP}\right\rangle\right|+\left|\beta_{UP}\right|^2\left|\left\langle p_V^{UP}|P|p_V^{UP}\right\rangle\right|}}
\end{equation}

Within the dipole and rotating wave approximations, the perturbation theory \cite{Akopian2006,Cohen-Tannoudji} gives for the two photon function 

\begin{equation}
\label{eq7}{
A_H^{LP}\equiv\alpha_{LP}\left\langle k_1,k_2|p_H^{LP}\right\rangle=\frac{x_{ex}^{H,LP} \sqrt{\Gamma_{XX}} x_{ph}^{H,LP} \sqrt{\Gamma_{H}^{LP}}/2\pi}{(\left|k_1\right|+\left|k_2\right|-\epsilon_{H}^{XX})(\left|k_2\right|-\epsilon_H^{LP})}
,}
\end{equation}
where $k_1$ and $k_2$ are the momenta of the photons and $\Gamma_{XX(LP)}$ is the line width of the biexciton (lower polariton). Furthermore, $\epsilon_{XX(LP)}=E_{XX(LP)}+i\Gamma_{XX(LP)}/2$ is the complex energy of the biexciton (lower polariton). The exciton (photon) Hopfield coefficients of the polariton state are denoted by $x_{ex(ph)}^{H,LP}$ and the polariton lifetime is given by the ratio of the square of the photon Hopfield coefficient and the cavity lifetime $\Gamma_{LP}=|x_{ph}^{H,LP}|^2/\tau_C$. A similar expression of Eq.(\ref{eq7}) can be obtained for the upper polariton state and the perpendicular polarization. The final equation for the off-diagonal element of the density matrix reads

\begin{equation}
\label{eq8}
\gamma'=\frac{\int{\int{dk_1 dk_2 A_H^{LP*} W A_V^{UP}}}}{\int{\int{dk_1 dk_2 A_H^{LP*} W A_H^{LP}}}+\int{\int{dk_1 dk_2 A_V^{UP*} W A_V^{UP}}}}.
\end{equation} 

The function $W$ corresponds to the spectral windows at the energies $E_{XX}$ and $E_{LP}^{H}$.

Finally, to estimate the quantum correlations of the emitted photons we use the Peres criterion for entanglement,\cite{Peres1996} which states that the emitted photons are entangled for $\gamma=1/2$ and not entangled for $\gamma=0$.

The density matrix of the system is in the so-called "x-form", containing only diagonal and anti-diagonal elements and thus another measure of entanglement -- the concurrence $C$ \cite{Wootters1998} -- is simply two times the absolute value of the off-diagonal element of the density matrix.\cite{Eberly,Yonac} 

The degree of entanglement is strongly correlated with the lineshape of the transitions as it follows from Eq.(\ref{eq7})and Eq.(\ref{eq8}): the better the overlap of the detected emission lines, the higher the off-diagonal element. The photoluminescence spectra for each transition can be calculated by integration of Eq.(\ref{eq7}) either over $k_2$ to obtain the biexciton-polariton emission line or over $k_1$ to obtain the polariton-ground state emission line. \cite{Cohen-Tannoudji} 

\begin{figure}
\begin{center}
\includegraphics[width=1.0\linewidth]{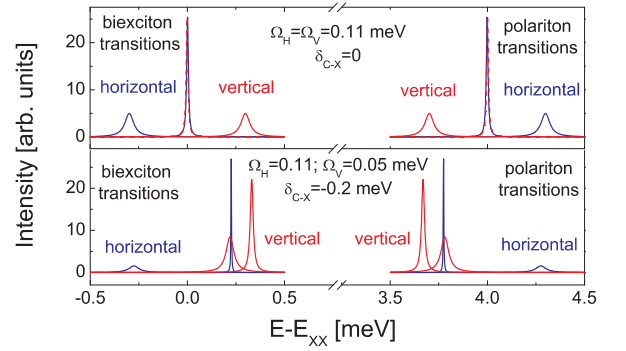}
\caption {
\label{fig2} Photoluminescence spectra for equal splittings (upper panel) at $\delta_{C-X}=0$ and for splittings $\Omega_H=0.11$ meV, $\Omega_V=0.05$ meV at $\delta_{X-C}=-0.2$ meV (lower panel). The blue(red) color corresponds to horizontal (vertical) polarization.} 
\end{center}
\end{figure}

Fig.\ref{fig2} shows the complete spectra of emission resulting from the biexciton decay. The upper panel is calculated for $\Omega_{H}=\Omega_{V}$=0.11 meV which corresponds to the experimentally measured value of \cite{Hennessy2006}
and the lower one for $\Omega_{H}$=0.11 meV and $\Omega_{V}$=0.05 meV. 
The complicated complete spectra shows 4 Lorentzian lines for the biexciton-polariton transitions with a linewidth $(\Gamma_{XX}+\Gamma_{P})$ and 4 Lorentzian lines with $\Gamma_{P}$ for the polariton-ground state transitions. The linewidth depend strongly on the photonic fraction of the polariton because the cavity photon lifetime is typically 100 times shorter that the
QD exciton lifetime. One should note that this type of spectra resulting from the biexciton decay in a strongly coupled microcavity has 
been recently measured but only for one polarization.\cite{Winger} One can see on the upper panel, that the resonance condition between the two polariton states H an V also corresponds to equal linewidth of the states and therefore to a high degree of entanglement $\gamma'=0.49$. On the other hand one can see on the lower panel of the figure that the nice symmetry of the scheme is broken when the oscillator strengths of the two resonances are different. The degree of entanglement is much lower $\gamma'=0.09$ in this last case, which we are going to analyze in details in the next section.

\section{Rabi splitting}

\begin{figure}
\begin{center}
\includegraphics[width=1.0\linewidth]{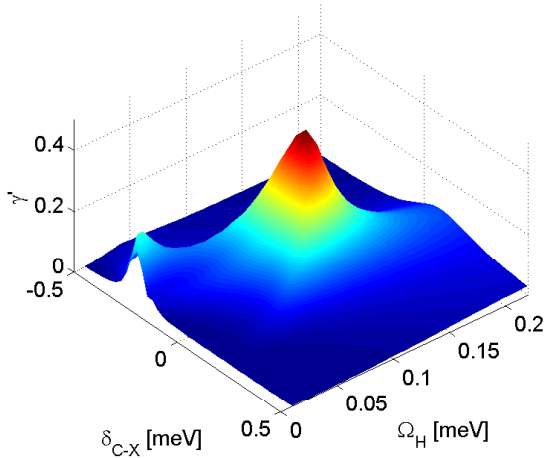}
\caption {
\label{fig3} Dependence of the off-diagonal element $\gamma'$ on the relative position of the cavity resonances and on the Rabi splitting for one polarization $\Omega_H$. The second splitting is kept constant at $\Omega_V=0.11$ meV.}
\end{center}
\end{figure} 

In the following we analyze the influence of different Rabi splittings of the H and V polariton states on the degree of entanglement $\gamma'$
of the two photon wave function resulting from the biexciton decay. This polarization anisotropy comes from the asymmetry of the QD or its environment and results in different oscillator strength for excitons which couple to different polarizations.\cite{Stier1999,Cantele2002,Favero2005,Santori2002} Also misalignment of the quantum dot and the antinode of the cavity mode field in one direction can change the coupling constant.\cite{Vuckovic} Consequently, the final polariton states of each polarization H and V have different values of Rabi splitting $\Omega_V$ and $\Omega_H$. 

Fig.\ref{fig3} shows the degree of entanglement depending on the relative position of the cavity resonances $\delta_{C-X}=(E_C^H+E_C^V)/2-(E_X^H+E_X^V)/2$ and on $\Omega_H$, whereas $\Omega_V$ is kept constant, equal to 0.11 meV.

\begin{figure}
\begin{center}
\includegraphics[width=1.0\linewidth]{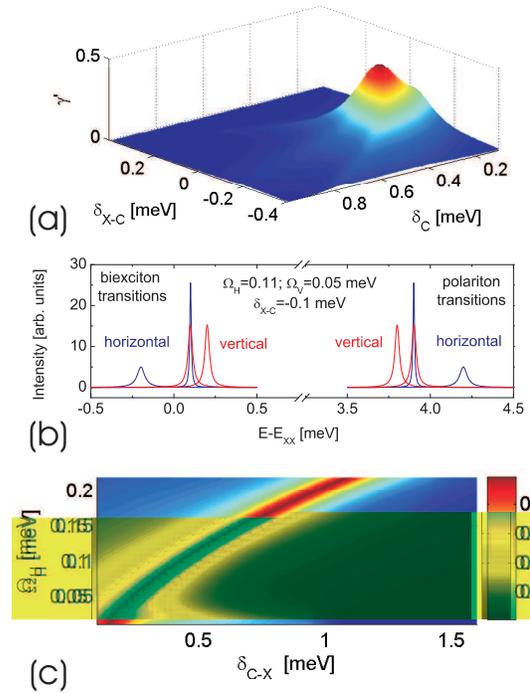}
\caption {
\label{fig4} (a) Dependence of the off-diagonal element $\gamma'$ on the relative position of the cavity resonances and on the splitting between the photonic modes $\delta_{C}$. The Rabi splittings are kept constant at $\Omega_H=0.11$ meV and $\Omega_V=0.05$ meV. (b) photoluminescence spectra for a splitting $\delta_{C}=0.3$ meV. The blue(red) color corresponds to horizontal (vertical) polarization. (c) Degree of entanglement versus cavity resonance splitting and Rabi splitting.} 
\end{center}
\end{figure} 

As one can see, the maximum degree of entanglement close to theoretical limit is achieved if both splittings are equal. If one changes the Rabi splitting of one polarization, the degree of entanglement decreases. In the case of $\Omega_H=\Omega_V$, the optimal detuning $\delta_{C-X}$ is zero. 
This optimal value of $\delta_{C-X}$ becomes positive when $\Omega_H-\Omega_V$ is positive and negative when $\Omega_H-\Omega_V$ is negative.
In addition to the main maximum, the degree of entanglement shows a three-peak structure for small $\Omega_H$ and a second small maximum. These effects arise from the interplay between the evolution of the lineshapes and overlaps of the transitions in the region where the polariton states are close to each other, depending on $\delta_{C-X}$. 

The difference of the Rabi splittings makes necessary a readjustment of the cavity resonances. In the following we are going to tune 
independently the energies of the H and V polarized modes. It means that now we tune not only $\delta_{C-X}$,
but also $\delta_{C}$. This can be achieved experimentally by atomic force microscope nano-oxidation of the cavity surface \cite{Hennessy2007}.
The results of the calculations are shown on Fig.\ref{fig4}.
We use $\Omega_H$=0.11 meV, $\Omega_V$=0.05 meV, and $\delta_{X}$=0.25 meV (as for the lower panel of figure \ref{fig2}), which represent a highly asymmetric case. Figure \ref{fig4} (a) shows  $\gamma'$ versus $\delta_{C-X}$ (x-axis) and $\delta_{C}$ (y-axis). One can see that the independent
tuning of the position of the two photon modes allows to recover quite a high value $\gamma'=0.41$. The photoluminescence spectra corresponding to this optimal configuration are shown on the figure\ref{fig4}(b). Once again, $\gamma'$ shows a complicated three-peak-structure when the detunings are not optimized (see Fig.\ref{fig4}(a)), which finds its origin in the interplay between the evolution of the lineshapes and overlaps of the transitions.

In Fig. \ref{fig4}(c), we keep $\Omega_V=0.11 meV$ constant and show the best
value of $\gamma'$ (which can be obtained tuning $\delta_{C-X}$) versus $\Omega_H$ and $\delta_C$. From this figure one can conclude that, whatever the value $\Omega_H-\Omega_V$, it is possible to find the values of $\delta_C$ and $\delta_{C-X}$ for which $\gamma'$ is larger than 0.4, which confirms the fact that
the detrimental effect induced by the difference of $\Omega_H$ and $\Omega_V$ can be, in all cases, overcome by the independent tuning of the cavity mode energies.

\section{Spectral filtering}

\begin{figure}
\begin{center}
\includegraphics[width=0.8\linewidth]{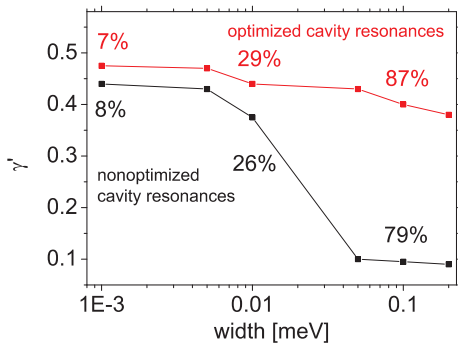}
\caption {
\label{fig5} Degree of entanglement versus the width of the spectral windows for non-optimized cavity resonances (black) and optimized cavity resonances (red). The percentages are the calculated quantum efficiencies.} 
\end{center}
\end{figure}

Spectral filters have been used by Akopian et al. in 2006 \cite{Akopian2006} to increase the quantum correlations of detected photons. Therein, the spectral windows were used to select the overlapping part of the non-degenerate exciton emission lines and thus a non-zero off-diagonal element of the density matrix has been observed. We define the quantum efficiency as the ratio between the number of photon pairs emitted and the number of photon pairs detected through the spectral window. This efficiency is expected to become smaller and smaller with the use of a sharper spectral window. The scheme we have proposed above already includes the use of spectral windows to preselect the appropriate transition lines, but there is no filtering in the sense that we always detect the whole emitting line. In the following, we analyze the impact of the width of the spectral windows on the entanglement degree and on the quantum efficiency of the biexciton decay. Fig.\ref{fig5} shows the dependence of $\gamma'$ on the width of the two identical windows around $E_p$ and $E_{XX}$ together with the evolution of the quantum efficiency. 
Two cases are considered, which 
are: the unfavorable case presented on the lower panel of the figure 2, and the one obtained for the same set of parameters but with the optimization of $\delta_{C}$ maximizing $\gamma'$. Both optimized and none optimized configurations show high $\gamma'$ and low quantum efficiency, below 10\% for the smallest width of spectral filters (<10$\mu eV$). Increasing this spectral width has a dramatic effect on the $\gamma'$
value for the non-optimized case. In contrast, the optimized setup allows to achieve a high degree of entanglement simultaneously with a large quantum efficiency by using a wide spectral window of 0.1 meV.

\section{Strongly coupled biexciton}

Another possibility which can be taken into account is the existence of additional resonances, e.g. a photonic mode close or at the biexciton-exciton transition energy. The idea is to accelerate the emission of entangled photon pairs by mixing the biexciton with a photon. We add now a V and an H-polarized photon fields at the biexciton-polariton transition energy with energies $E_{C_{XX}}^H$ and $E_{C_{XX}}^V $. One should therefore have a reversible coupling between three possible configurations for each polarization which are: the biexciton, one exciton and one photon, and two photons. Taking into account the polarization degree of freedom and the fact that the bi-exciton is a common state for the two polarization channels, we should describe the reversible coupling between 5 states. The eigenenergies of the resulting dressed states can be found by the diagonalization of the 5x5 matrix $M$, where $\Omega_{H(V)}^{XX}$ is the coupling to the biexciton state:

\begin{widetext}
\begin{equation}
\label{eq9} { 
M=  \begin{pmatrix} E_{XX} & \Omega_H^{XX} & \Omega_V^{XX} & 0 & 0 \\ \Omega_H^{XX} & E_X^H+E_{C_{XX}}^H & 0 & \Omega_H & 0 \\ \Omega_V^{XX} & 0 & E_X^V+E_{C_{XX}}^V & 0 & \Omega_V \\ 0 & \Omega_H & 0 & E_C^H+E_{C{XX}}^H & 0 \\ 0 & 0 & \Omega_V & 0 & E_C^V+E_{C{XX}}^V
\end{pmatrix}.
}
\end{equation}
\end{widetext}

The structure of the five eigenvalues and eigenvectors of this matrix is quite complicated in the general case. It is possible to tune $E_{C_{XX}}^{H}$ and $E_{C_{XX}}^{H}$ in such a way that the two polarized biexciton-polariton transitions from one initial bipolariton state are symmetric, which means that both polarization paths have the same radiative lifetime and thus the same linewidth. This results in a high degree of entanglement if we assume that the polariton-ground state transitions are optimized and unperturbed. In addition the decay from a "photon-like" bipolariton state is much faster in comparison to the uncoupled biexciton. The full transition to the ground state will take place in a few ps and repetition rates close to THz range become realistic.

However, the adjustment of four different photonic resonances at the same time seems to be extremely challenging. Furthermore the resulting fine structure of the bipolariton is complicated and the selection of the transition lines would be also a difficult problem, which would end up by a huge reduction of the quantum efficiency. Thus, we do not believe at this stage that a  biexciton strongly coupled to light modes in addition to the strong coupling of the excitons would be really advantageous for applied purposes. 

Nevertheless, a photonic resonance weakly coupled to the biexciton-polariton transition may accelerate the first photon emission, either by Purcell effect, or simply by reducing the quenching of the emission which could be provoked by the fact that the resonance is placed within a photonic bandgap. Also dressing the biexciton state modifies the transition properties \cite{Jundt2008, Muller2008} and stays as a tool for future applications.

\section{Summary and conclusions}

We have shown that the difference of exciton oscillator strengths of the exciton states coupled to V and H polarized light strongly affects the entanglement degree of the photon pairs emitted during the biexciton decay, when the excitonic resonance of the QD is strongly coupled to the cavity modes of a photonic crystal. However, we have shown that this detrimental effect can be compensated if it is possible to tune independently the energies
of the polarized photonic modes, which has been demonstrated to be possible experimentally.\cite{Hennessy2007} We have also analyzed the impact of a spectral filtering of the different emission lines, showing that an increase of the entanglement degree by this method has to be paid for by a strong reduction of the quantum efficiency. Finally, we have discussed the possible impact of the presence of a cavity mode resonant with the biexciton transition. We found that this coupling can indeed accelerate the biexciton decay and give access to very high repetition rates (close of one THz) for the entangled photon emission. However, the complication brought by the presence of many polariton lines which should be all tuned simultaneously makes this configuration very hard to implement for an experimental point of view. We conclude that the control of the electronic resonances through their strong coupling to confined cavity modes opens new perspectives and is from many points of view extremely advantageous for the fabrication of a solid source of entangled photon pairs emitted on demand.

The authors thank D. Solnyshkov and I. Shelykh for useful discussions and critical reading of the manuscript. The work was supported by the ANR Chair of Excellence and EU FP6-517769 STIMSCAT.

\end{document}